\documentstyle[12pt,epsf,epsfig]{article}
\def\unit{\hbox to 3.3pt{\hskip1.3pt \vrule height 7pt width .4pt \hskip.7pt
\vrule height 7.85pt width .4pt \kern-2.4pt
\hrulefill \kern-3pt
\raise 4pt\hbox{\char'40}}}

\newcommand{\gtrsim}
{\ \rlap{\raise 2pt\hbox{$>$}}{\lower 2pt \hbox{$\sim$}}\ }
\newcommand{\lessim}
{\ \rlap{\raise 2pt\hbox{$<$}}{\lower 2pt \hbox{$\sim$}}\ }
 
\newcommand{\ea}{{ et al.}}

\def\identity{1 \hspace{-.085cm}{\rm l}}

\def\beq{\begin{equation}}
\def\eeq{\end{equation}}
\def\bea{\begin{eqnarray}}
\def\eea{\end{eqnarray}}
\def\ba{\begin{array}}
\def\ea{\end{array}}
\def\gappeq{\mathrel{\rlap {\raise.5ex\hbox{$>$}}
{\lower.5ex\hbox{$\sim$}}}}
\def\permil{$\%\raise.20ex\hbox{$_0$}}
\def\lappeq{\mathrel{\rlap{\raise.5ex\hbox{$<$}}
{\lower.5ex\hbox{$\sim$}}}}
\begin{document}
\topmargin -1.0cm
\oddsidemargin -0.8cm
\evensidemargin -0.8cm
\pagestyle{empty}
\begin{flushright}
CERN-TH/96-255
\end{flushright}
\vspace*{5mm}
\begin{center}
{\Large \bf  Multi-Messenger Theories of Gauge-Mediated  
Supersymmetry Breaking}\\
\vspace{2.5cm}
{\large S.~Dimopoulos}\\
\vspace{0.3cm}
{\small Theory Division, CERN,
Geneva, Switzerland\\
and\\
Physics Department, Stanford University, USA}\\
\vspace{0.5cm}
{\large G.F.~Giudice\footnote{On leave of absence from INFN, Sezione di
Padova,
Padua, Italy.}}\\ 
\vspace{0.3cm}
{\small Theory Division, CERN,
Geneva, Switzerland}\\
\vspace*{2cm}
Abstract
\end{center}

We study gauge-mediated theories containing several messengers with the 
most general SU(5)-invariant mass and supersymmetry-breaking parameters. 
We show that these theories are predictive, containing only two relevant 
parameters more than the minimal gauge-mediated model. Hypercharge D-terms 
can contribute significantly to the right-handed charged sleptons and bring 
them closer in mass to the left-handed sleptons. The messenger masses must 
be invariant under either SU(5) or a ``messenger parity" to avoid 
spontaneous breaking of charge conservation. 
 
\vfill
\begin{flushleft}
CERN-TH/96-255\\
September 1996
\end{flushleft}
\eject
\pagestyle{empty}
\setcounter{page}{1}
\setcounter{footnote}{0}

\baselineskip20pt
\pagestyle{plain}

Theories in which supersymmetry breaking is 
communicated to the observable fields 
through gauge interactions \cite{gau}, instead of gravity,
have the advantage of insuring a correct suppression of flavour-changing
neutral currents. Another interesting feature of these theories is their
sharp predictivity of the new particle mass spectrum. Let us first
consider the most
restrictive case \cite{din}
in which the messenger fields responsible for the 
communication of supersymmetry breaking form a single 
representation of a GUT model. Let us also define $M$ as the mass of the
messenger supermultiplet and $\sqrt{F}$ as the supersymmetry-breaking 
scale, or, in other words, as the mass splitting inside the messenger
supermultiplet. In this minimal case, the complete mass spectrum of squarks,
sleptons, and gluinos is described only by the parameter $F/M$, beside a mild
logarithmic dependence on $M$. 

A complication arises in the Higgs
sector, as the Higgs mass parameters which violate a Peccei-Quinn symmetry
cannot be generated by gauge interactions alone. In order to parametrize
some new unknown interactions, two free inputs
have to be introduced, denoted by $\mu$ and $B_\mu$ with standard conventions.
One of these two parameters is determined by the electroweak symmetry breaking
condition. Notice however that if $\mu$ and $B_\mu$ are generated radiatively
by some new interaction, then it is fairly generic \cite{noi}
to obtain new contributions to the Higgs mass parameters, usually denoted
by $m_{H_1}^2$ and $m_{H_1}^2$. Thus, unless one relies on specific models
for the Peccei-Quinn violating interactions, the Higgs sector introduces
three new free parameters to the theory. However the predictions 
for slepton, squark, and gluino masses do not depend on these parameters.

As these mass predictions may soon face experimental test, it is important
to establish how much they depend on the minimality of the model or how
much they descend from the gauge-mediation principle. Variations of the
minimal model have already been considered in refs.~\cite{var,mar}.
Direct superpotential couplings between messenger and observable fields
have been analysed in ref.~\cite{var}. Although these couplings spoil
the natural flavour conservation of gauge-mediated supersymmetry-breaking
theories, they have the advantage of insuring a fast decay of the lightest
messenger, thus avoiding cosmological problems \cite{dar}. Because of
an accidental cancellation\footnote{This cancellation was first pointed 
out in ref.~\cite{noi}, in the case of couplings between Higgs and
messenger superfields.}, the new flavour-violating
contributions to squark and slepton masses are suppressed by the ratio
$F^2/M^4$ and 
are therefore less dangerous than one may have naively expected \cite{var}.
The author of ref.~\cite{mar} has investigated the case in which the
messenger fields do not form complete GUT representations. Although the
successful gauge-coupling unification is usually lost, it
is interesting to know that several of the mass predictions are approximately
preserved, despite the large number of discrete choices for the messenger
representations.

In this paper we want to study a generalization of the minimal model.
We will consider the case in which the messengers form $n$ copies of the same
real GUT representation and their supersymmetric and supersymmetry-breaking
mass matrices have a completely generic structure invariant under GUT
symmetry.
For definiteness, we will take $n$ ${\bf 5}+{\bf
\overline 5}$ of $SU(5)$, called $\Phi_i$ and $\bar \Phi_i$ ($i=1,...,n$), 
and the extension of our results to
other representations is completely straightforward. 
The messenger mass matrix is defined by the superpotential term
\beq
W={\bar \Phi}_i M_{ij}^\Phi \Phi_j ~~~~~~~~i,j=1,...,n
\label{uno}
\eeq
and by a supersymmetry-breaking term in the scalar potential
\beq
V={\bar \Phi}_{i} F_{ij}^\Phi \Phi_{j}+{\rm h.c.}
\label{due} 
\eeq
Here $\Phi_i$ label either the $SU(2)$ doublet or the $SU(3)$
triplet components of
the $i$-th ${\bf 5}$-plet messenger. 
After GUT-symmetry spontaneous breaking, the mass parameters for
the doublets and the triplets, distiguished by
the index $\Phi$ on $M^\Phi$ and $F^\Phi$ in eqs.~(\ref{uno})--(\ref{due}),
renormalize differently.
Our basic hypothesis is that the mass scales $M^\Phi$ and $F^\Phi$ are generated by 
a sector neutral under GUT interactions, and therefore at the
GUT scale 
$M^\Phi$ and $F^\Phi$ are the same for both doublets and triplets.
We will refer to this as the ``GUT singlet hypothesis".
If we define $M$ and $F$ as the common
values of the mass matrices
at the GUT scale $M_{GUT}$, a one-loop renormalization scaling gives the values
of the different $M^\Phi$ and $F^\Phi$ at the energy scale $Q$:
\beq
\frac{F^\Phi(Q)}{F}=\frac{M^\Phi(Q)}{M}=
\prod_{r=1}^3\left[ \frac{\alpha_r (Q)}{\alpha_r(M_{GUT})}
\right]^{-2\frac{C^\Phi_r}{b_r}}~.
\label{run}
\eeq
Here $C^\Phi$ depends on the Standard Model quantum numbers,
$C=\frac{N^2-1}{2N}$ for the $N$-dimensional representation of
$SU_N$, and $C=Y^2$ ($Y=Q-T_3$) for the $U_1$ factor. Also 
$b_r$ are the $\beta$-function
coefficients
\beq
b_3=-3+n,~~~ b_2=1+n,~~~ b_1=11+\frac{5}{3}n,
\label{bis}
\eeq
and $n$ counts the messenger 
contribution. Equation (\ref{run}) shows that the ratio 
$F^\Phi(Q)/M^\Phi(Q)$ is independent of the energy scale $Q$. The
``GUT singlet hypothesis" implies also that this ratio is independent
of $\Phi$, {\it i.e.} it is equal for the doublet and triplet
messenger components, $F^\Phi(Q)/M^\Phi(Q)=F/M$. 

We will take here $M_{ij}$ and $F_{ij}$ as general $n\times n$ matrices,
because of our lack of knowledge on the sector which originally breaks
supersymmetry. A simpler case can be considered,
in which
both $M_{ij}$ and $F_{ij}$ originate from
the couplings of messengers to a single superfield which acquires vacuum 
expectation values in the scalar and auxiliary components. Then $M_{ij}$
and $F_{ij}$ are proportional, and the model simply reduces to $n$
replicas of the minimal model considered above. 
We believe that there are no strong reasons to make the restrictive
assumption that $M_{ij}$ and $F_{ij}$ are proportional. Moreover
we will show that
the case of generic $M_{ij}$ and $F_{ij}$ has a richer structure
and leads to important differences 
in the physical mass spectrum.

Let us start analyzing the supersymmetry-breaking masses in the
observable sector induced by radiative corrections.
After a redefinition
of the messenger superfields, we can choose, without loss of generality, 
$M_{ij}$ to be diagonal with
real and positive eigenvalues $M_i$ ($i=1,...n$). The gaugino masses
arise at one loop
\beq
m_{\lambda_r}=k_r ~\alpha_r \frac{\Lambda_G}{4\pi} ~\left[ 
1+{\cal O}(F^2/M^4) \right]
~,~~~~~~~r=1,2,3~,
\label{gaug}
\eeq
where $k_1=5/3$, $k_2=k_3=1$, the gauge coupling constants are normalized
such that $k_r\alpha_r$ ($r=1,2,3$)
are all equal at the GUT scale, and 
\beq
\Lambda_G=\sum_{i=1}^n\frac{F_{ii}}{M_i}~.
\label{lamg}
\eeq
Here and in the following, we use the approximation that all entries of
the $F$ matrix are smaller than the entries of $M^2$. This is justified
because some of the scalar messenger square masses can become negative if some
eigenvalues of $F$ are larger then the corresponding one in $M^2$. The
deviations from the leading-order expansion are generally small \cite{mar,dar}
unless $F/M^2$ is extremely close to 1. Notice also that $\Lambda_G$
does not depend on the energy scale at which is defined, as a consequence
of the non-renormalization of the ratio $F/M$, see eq.~(\ref{run}).
In eq.~(\ref{gaug}) $\alpha_r$ is evaluated at the scale $M$. The one-loop
renormalization group running of $m_{\lambda_r}$ just amounts to
using eq.~(\ref{gaug}) with $\alpha_r$ evaluated at $m_{\lambda_r}$.

The two-loop contributions to squark and slepton masses are given by
\beq
m_{\tilde f}^2=2\sum_{r=1}^3 C^{\tilde f}_r~
k_r~\alpha_r^2 \left( \frac{\Lambda_S}{4\pi}\right)^2
~\left[ 1+{\cal O}(F^2/M^4) \right]~.
\label{squak}
\eeq
$C^f$ depends on the quantum number of squarks or sleptons as explained
after eq.~(\ref{run}). 
In the limit
in which $M$ is proportional to the identity
($M=M_0\times \identity$), $\Lambda_S$ is given by
\beq
\Lambda_S=\left(\sum_{i,j=1}^n\frac{|F_{ij}|^2}{M_0^2}\right)^{1/2}~.
\label{lams}
\eeq
In models where the matrices $M$ and $F$ are both
proportional to the identity, $\Lambda_G/\Lambda_S =\sqrt{n}$, but,
in the general case,
$\Lambda_G$ and $\Lambda_S$ are independent values.
Eq.~(\ref{squak}) is valid at the scale $M$. Throughout
the paper we follow the convention that $\alpha_r$ is always
evaluated at $M$, unless
indicated otherwise. After the appropriate
one-loop renormalization running, $m_{\tilde f}^2$ becomes
\beq
m_{\tilde f}^2=2\sum_{r=1}^3 C^{\tilde f}_r~
k_r~\alpha_r^2 ~
\left[ \left( \frac{\Lambda_S}{4\pi}\right)^2 + 
\frac{k_r(1-\xi^{2}_r)}{b_r}\left( \frac{\Lambda_G}{4\pi}\right)^2  \right]~,
\label{squakr}
\eeq
where $\xi_r\equiv \alpha_r (m_{\tilde f})/\alpha_r$ and $b_r$
are given in eq.~(\ref{bis}) taking $n=0$.

It is known that squark and slepton masses
can also be generated at one
loop, but these contributions are proportional to hypercharge and therefore
not positive definite. It is usually assumed that, in a realistic model,
such contributions have to vanish \cite{din}. In the theories under 
consideration, they are given by
\beq
\Delta m_{\tilde f}^2=\frac{\alpha_1}{4\pi}Y_{\tilde f}~ {\rm Tr} Y_\Phi ~
 {\bar \Lambda}^{\Phi 2}_D ~,
\label{1loo}
\eeq
where, at the leading order in $F/M^2$, ${\bar \Lambda}^{\Phi 2}_D$ is
independent of the specific component of the GUT multiplet $\Phi$,
\beq
{\bar \Lambda}^{\Phi 2}_D \equiv
{\bar \Lambda}^2_D = \frac{1}{2}
\sum_{i,j=1}^n \frac{|F_{ji}|^2-|F_{ij}|^2}{M_i^2}f\left(
\frac{M_j^2}{M_i^2}\right)~,
\label{ld}
\eeq
\beq
f(x)=\frac{2}{(1-x)}+\frac{(1+x)}{(1-x)^2}\ln x~.
\label{fx}
\eeq
For simplicity, in the following 
we will focus mainly on the case $n=2$,
in which eq.~(\ref{ld}) reduces to
\beq
{\bar \Lambda}_D^2=
\frac{|F_{21}|^2-|F_{12}|^2}{M_1^2}f\left(
\frac{M_2^2}{M_1^2}\right)~.
\eeq
As proved in ref.~\cite{noi}, the hypercharge D-term contributions to scalar
masses vanish up to two loops, if $F$ is hermitian in the basis in which
$M$ diagonal, real, and positive. This is a consequence a
symmetry transforming the messenger superfields $\Phi$, $\bar \Phi$, and
the gauge vector superfields $V$ as follows \cite{noi}
\beq
\Phi \to U {\bar \Phi}^\dagger ~, ~~~~
{\bar \Phi} \to \Phi^\dagger  {\bar U}~, ~~~~
V \to -V~,
\eeq
where $U$ and $\bar U$ are arbitrary $n\times n$ unitary matrices. The messenger
sector is invariant under this ``messenger parity" transformation
provided that, independently of the specific basis, one can find some
matrices $U$ and $\bar U$ such that
\beq
M^\dagger ={\bar U}MU ~,~~~F^\dagger ={\bar U}FU ~.
\eeq
As ``messenger parity" is broken by ordinary particles, hypercharge D-terms
mass contributions can arise at higher-loop order, but these are too
small to play any significant role in the supersymmetric particle mass
spectrum.
As a consequence of ``messenger parity" eq.~(\ref{ld}) vanishes
when $F_{ij}=F_{ji}^\star$. Notice also that eq.~(\ref{ld}) vanishes
if $M$ is proportional to the identity. Indeed if $M\propto \identity$,
we can always rotate the messenger scalar fields and make $F$ hermitian.

In the absence of ``messenger parity", eq.~(\ref{1loo}) in general leads
to negative square masses for either ${\tilde e}_L$ or ${\tilde e}_R$.
However if the ``GUT singlet hypothesis" is valid, then
eq.~(\ref{1loo}) vanishes simply because ${\rm Tr}Y_\Phi =0$ \cite{noi}. 
This cancellation is
guaranteed by the universality of
${\bar \Lambda}^\Phi_D$ within the GUT multiplet, or, in other words, by the 
non-renormalization of the ratio $F/M$. It is then clear that this
cancellation will not persist to higher orders in $F/M^2$. At next order,
${\bar \Lambda}^\Phi_D$ 
depends on the different Standard Model quantum numbers of
the messenger fields, and therefore it acts non-trivially under the 
trace in eq.~(\ref{1loo}):
\beq
{\bar \Lambda}_D^{\Phi 2} =\sum_{i,j,k,l}\left[ (F^{\Phi 
\dagger})_{ij}F^\Phi_{jk}(F^{\Phi \dagger})_{kl}
F^\Phi_{li}-F^\Phi_{ij}(F^{\Phi \dagger})_{jk}F^\Phi_{kl}(F^{\Phi 
\dagger})_{li}\right] T_{ijkl}~,
\label{ord}
\eeq
\beq
T_{ijkl}=\frac{i}{\pi^2}\int d^4k \frac{1}{(k^2-M^{\Phi 2}_i)^2}
\frac{1}{k^2-M^{\Phi 2}_j}\frac{1}{k^2-M^{\Phi 2}_k}
\frac{1}{k^2-M^{\Phi 2}_l}~.
\eeq
In the case $n=2$, eq.~(\ref{ord}) reduces to
\begin{eqnarray}
{\bar \Lambda}_D^{\Phi 2} 
&=&\frac{|F^{\Phi}_{12}|^2-|F^{\Phi}_{21}|^2}{M^{\Phi 6}_1}
\left[ |F^\Phi_{11}|^2 ~g_1\left(\frac{M^{\Phi 2}_2}{M^{\Phi 2}_1}\right)-
|F^\Phi_{22}|^2 ~ \frac{M^{\Phi 6}_1}{M^{\Phi 6}_2}~
g_1\left(\frac{M^{\Phi 2}_1}{M^{\Phi 2}_2}\right)\right. \nonumber \\
&+&\left. \left( |F^\Phi_{12}|^2 +|F^\Phi_{21}|^2\right) ~
g_2\left(\frac{M^{\Phi 2}_2}{M^{\Phi 2}_1}\right)\right]~,
\label{ord2}
\end{eqnarray}
\beq
g_1(x)=\frac{x^2-8x-17}{6(1-x)^3}-\frac{1+3x}{(1-x)^4}\ln x ~,
\eeq
\beq
g_2(x)=\frac{x^2+10x+1}{2x(1-x)^3}+\frac{3(1+x)}{(1-x)^4}\ln x~.
\eeq
With the help of the expressions for $M^\Phi$ and $F^\Phi$ 
in terms of their boundary conditions
at the GUT scale $M$ and $F$, see eq.~(\ref{run}), we can explicitly
evaluate the trace in eq.~(\ref{1loo}):
\beq
{\rm Tr} Y_\Phi {\bar \Lambda}^{\Phi 2}_D = \left[ 
\left(\frac{\alpha_3}{\alpha_X}\right)^{\frac{16}{3b_3}}
\left(\frac{k_1~\alpha_1}{\alpha_X}\right)^{\frac{4}{9b_1}}-
\left(\frac{\alpha_2}{\alpha_X}\right)^{\frac{3}{b_2}}
\left(\frac{k_1~\alpha_1}{\alpha_X}\right)^{\frac{1}{b_1}}\right] 
{\bar \Lambda}_D^{(GUT)2}~.
\label{sum}
\eeq
Here ${\bar \Lambda}^{(GUT)2}_D$ 
is given by eq.~(\ref{ord})
evaluated at the GUT energy scale or, in other words, by these equations
with the index $\Phi$ suppressed. 
Also $\alpha_X=k_r\alpha_r(M_{GUT})$ for any $r=1,2,3$.

For $n=2$ eq.~(\ref{sum}) gives ${\rm Tr}Y_\Phi {\bar \Lambda}_D^{\Phi 2} 
\simeq -0.3 {\bar \Lambda}^{(GUT)2}_D$. 
Nevertheless ${\bar \Lambda}_D^{(GUT)2}$ can be quite smaller
than $\Lambda_S^2$, because of
the extra
$F^2/M^4$ factor. Actually,
if $\sqrt{F}\lappeq (\alpha_3 / \pi )^{1/4} M \simeq 0.4 M$, the most
important contribution from the hypercharge D-term comes at the two-loop
order, from the diagrams shown in fig.~1. Evaluation of these diagrams,
for the case $n=2$, gives
\beq
{\bar \Lambda}^{\Phi 2}_D=\sum_{r=1}^3 \frac{\alpha_r}{\pi}C^\Phi_r
\frac{|F_{21}|^2-|F_{12}|^2}{M_1^2}f\left(
\frac{M_2^2}{M_1^2}\right)~,
\eeq
where the function $f(M_2^2/M_1^2)$ is given in eq.~(\ref{fx}).
The trace over the fundamental $SU(5)$ messenger representation gives
\beq
{\rm Tr} Y_\Phi {\bar \Lambda}^{\Phi 2}_D = \left( \frac{4}{3}
\frac{\alpha_3}{\pi} -
\frac{3}{4}\frac{\alpha_2}{\pi}-\frac{5}{36}\frac{\alpha_1}{\pi} \right)
\frac{|F_{21}|^2-|F_{12}|^2}{M_1^2}f\left(
\frac{M_2^2}{M_1^2}\right)~.
\eeq
The two-loop contribution to squark and slepton square masses
from the hypercharge D-term can then be written as
\beq
\Delta m_{\tilde f}^2=Y_{\tilde f}\alpha_1
\left( \frac{16}{3}\alpha_3 -
3\alpha_2-\frac{5}{9}\alpha_1 \right) \left(
\frac{\Lambda_D}{4\pi}\right)^2 ~,
\label{mas}
\eeq
\beq
\Lambda_D^2\equiv
\frac{|F_{21}|^2-|F_{12}|^2}{M_1^2}f\left(
\frac{M_2^2}{M_1^2}\right)~.
\label{lad}
\eeq

For definiteness let us choose $M_1<M_2$. 
The function $f(M_2^2/M_1^2)$, given in eq.~(\ref{fx}), varies between
0 and about 0.1. Therefore, we expect that $\Lambda^2_D/\Lambda^2_S$
can be at most equal to about 0.1. The effect of $\Lambda_D$ is therefore
rather small for squark masses. However, in the case of
sleptons, the $\alpha_3$ present in eq.~(\ref{mas}) can compensate the
smallness of the
ratio $\Lambda^2_D/\Lambda^2_S$, and the hypercharge D-term contribution
can be of the same order of the usual gauge contribution.

As an example, we have plotted in fig.~2, as functions of
$\Lambda^2_D/\Lambda^2_S$, the masses 
of the left-handed and right-handed selectron, derived from
eqs.~(\ref{squakr}) and (\ref{mas}).
For simplicity we ignore 
the tree-level contribution of the D terms coming from
electroweak symmetry breaking, which are generally quite small.
We have fixed $\Lambda_S$ such that $m_{\tilde e_R}=100$ GeV if
$\Lambda_D=0$, and both selectron masses scale linearly with $\Lambda_S$.
We have chosen $\alpha_3=0.08$ at the messenger scale $M$, and
$\Lambda_G=\Lambda_S$. Results depend only very weakly on the choice
of $\Lambda_G$, unless $\Lambda_G\gg \Lambda_S$. 
The effect of non-vanishing $\Lambda_D$ is important for
$m_{\tilde e_R}$.
For $\Lambda^2_D/\Lambda^2_S=0.1$  the ratio $m_{\tilde e_L}/
m_{\tilde e_R}$ is 1.4,
instead of its value of 2.1 at $\Lambda_D=0$. This leads to similar  
production cross sections for $\tilde e_L$ and $\tilde e_R$
at the Tevatron and it  
could be important for the interpretation of the $e e \gamma \gamma
 {E\llap{/}}_T$ event  
reported by the CDF collaboration
\cite{cdf,ddt}.
Notice that the sign of $\Lambda^2_D$ is
not determined, and $m_{\tilde e_R}$ is
substantially reduced if $\Lambda^2_D/\Lambda^2_S$ is negative. Indeed
large negative values of
$\Lambda^2_D/\Lambda^2_S$ can be excluded by the experimental lower
bound on the selectron mass.

The new mass contribution from hypercharge D term will also affect
the Higgs mass parameters $m_{H_1}^2$ and $m_{H_2}^2$ at the scale $M$:
\beq
m_{H_1}^2=
\left( \frac{3}{2}\alpha_2^2+\frac{5}{6}\alpha_1^2\right)
\left( \frac{\Lambda_S}{4\pi}\right)^2 - \alpha_1
\left( \frac{8}{3}\alpha_3-\frac{3}{2}\alpha_2-
\frac{5}{18}\alpha_1\right)
\left( \frac{\Lambda_D}{4\pi}\right)^2 ~,
\label{h1}
\eeq
\beq
m_{H_2}^2=
\left( \frac{3}{2}\alpha_2^2+\frac{5}{6}\alpha_1^2\right)
\left( \frac{\Lambda_S}{4\pi}\right)^2 + \alpha_1
\left( \frac{8}{3}\alpha_3-\frac{3}{2}\alpha_2-
\frac{5}{18}\alpha_1\right)
\left( \frac{\Lambda_D}{4\pi}\right)^2 ~.
\label{h2}
\eeq
As mentioned above, any radiative mechanism which generates the Peccei-Quinn
violating terms $\mu$ and $B_\mu$ will also give new contributions to
$m_{H_1}^2$ and $m_{H_2}^2$ \cite{noi}. We assume 
here for simplicity that these
contributions are negligible with respect to those in 
eqs.~(\ref{h1})--(\ref{h2}). 

The running below the messenger scale $M$ does not considerably modify
$m_{H_1}^2$ but, because of the stop and gluino contributions, has
a very important effect on $m_{H_2}^2$. Keeping only the leading
terms proportional to the strong gauge coupling constant and to the
top Yukawa coupling, the effect of running is to induce an extra
contribution to eq.~(\ref{h2}):
\beq
\delta m_{H_2}^2=-\alpha_3^2\frac{\alpha_h t}{\pi}
\left\{ \left[ \frac{\alpha_3 t}{\pi}\xi_3
-\frac{9}{8}\frac{\alpha_h t}{\pi}\left( 1- I(\xi_3 )\right)^2\right]
\left( \frac{\Lambda_G}{4\pi}\right)^2+4I(\xi_3 )\left(
\frac{\Lambda_S}{4\pi}\right)^2\right\} ~,
\label{dh2}
\eeq
\beq
I(\xi_3 )\equiv \frac{9}{7}\left( \frac{1-\xi_3^{-7/9}}{\xi_3 -1}\right)
~~~~~~
\xi_3 \equiv \frac{\alpha_3({\tilde m})}{\alpha_3}=\left( 1-\frac{
3\alpha_3}{4\pi}t\right)^{-1}~.
\eeq
Here $t=\ln (M^2/{\tilde m}^2)$ where $\tilde m$ is the typical stop
(or gluino) mass scale; $\alpha_h=h_t^2/(4\pi)$, and $h_t$ is the 
top-quark Yukawa coupling at the scale $\tilde m$ 
related to the running top-quark mass
$m_t$ and to the ratio of Higgs vacuum expectation values 
$\langle H_2 \rangle /\langle H_1 \rangle =\tan \beta$ by the equation
\beq
h_t=\left( \frac{m_t}{174~{\rm GeV}}\right) \frac{1}{\sin \beta} ~.
\eeq

The electroweak symmetry breaking condition determines the parameter
$\mu$ in terms of $m_{H_1}^2$, $m_{H_2}^2$, and $\tan \beta$
\beq
\mu^2=\frac{m_{H_1}^2-m_{H_2}^2\tan^2\beta}{\tan^2\beta -1}-
\frac{M_Z^2}{2}~.
\label{mu}
\eeq
Using eqs.~(\ref{h1}), (\ref{h2}), and (\ref{dh2}), we can now express
$\mu$ as a function of $\Lambda_S$, $\Lambda_G$ and $\Lambda_D$. 
The hypercharge D term has the effect to reduce or increase the
value of $\mu$, if $\Lambda_D^2$ is positive or negative respectively.
A reduction in $\mu$ is very welcome, as
it reduces the amount of fine tuning of the theory \cite{nat}. Indeed, in gauge-mediated
supersymmetry breaking models, the steep running of $m_{H_2}^2$ in the
electroweak symmetry breaking condition, see
eq.~(\ref{dh2}), can only be compensated by a large $\mu$.
Therefore the physical value of $M_Z$ is obtained at the price of a
rather accurate cancellation between the term proportional to $\Lambda_S^2$
and $\mu^2$. The hypercharge D term can however reduce $\mu$ only up
to 5\%, if $\Lambda_D^2/\Lambda_S^2<0.1$. The effect is therefore too small
to solve the fine-tuning problem. On the other hand, the hypercharge
D term can increase the 
value of $m_{\tilde e_R}$, allowing
lower values of $\Lambda_S$ consistent with experimental limits. In this
respect it is possible, for $\Lambda_D^2/\Lambda_S^2=0.1$ and
for a fixed $m_{\tilde e_R}$, to obtain
values of $\mu$ 30--40\% smaller than for $\Lambda_D=0$, and therefore
alleviate the fine-tuning problem.

In conclusion we have studied theories in which supersymmetry breaking 
is mediated by
a multi-messenger sector with GUT-invariant mass parameters.
The mass spectrum of the supersymmetric particles is described by
three parameters $\Lambda_G$, $\Lambda_S$, and $\Lambda_D$. These
three mass scales are in general independent, but all of the same
order of magnitude $\cal{O}$(100 TeV). The relations between the 
physical masses and the scales $\Lambda_{G,S,D}$ are completely
specified by the particle gauge quantum numbers. $\Lambda_G$  determines
the gaugino masses, while $\Lambda_S$ contributes to the squark and
slepton masses. $\Lambda_D$ corresponds
to the hypercharge D-term contribution to slepton and squark masses,
and has no analogue in the single-messenger
case. 
It generates
scalar masses at one-loop, but its contribution is
suppressed by a factor $F^2/M^4$. It also gives a two-loop contribution,
which can be of the same order of magnitude as the $\Lambda_S$ contribution
for the right-handed charged sleptons, and smaller for the other
supersymmetric particles. Therefore the minimal model mass
relations are robust among supersymmetric particles with the same spin, 
with the exception of the right-handed slepton masses which are a good 
measure of possible hypercharge D-term contributions.
These conclusions rely on the
assumption that the strong dynamics responsible for
messenger masses is invariant under the SU(5) symmetry or, in other
words, that $M$ and $F$ 
are the same for the doublet and triplet components of the messenger
five-plet at $M_{GUT}$. If this is not the case, the messenger sector
must have a new ``parity" symmetry, which forbids dangerous 
mass square contributions to scalar particles at one loop. This
symmetry then guarantees the same mass relations among supersymmetric
particles with the same spin as in the minimal model, while scalar
and fermion masses depend on two different parameters.

We thank A.~Pomarol for important discussions and for participating
at the early stage of this work.
We are grateful to G.~Degrassi, S.~Fanchiotti, and P.~Gambino for 
permission to use
the program ProcessDiagram. We also thank A.~Vicini for help with 
ProcessDiagram and very useful discussions, and S.~Thomas and J.~Wells
for correspondence.

\def\ijmp#1#2#3{{\it Int. Jour. Mod. Phys. }{\bf #1~}(19#2)~#3}
\def\pl#1#2#3{{\it Phys. Lett. }{\bf B#1~}(19#2)~#3}
\def\zp#1#2#3{{\it Z. Phys. }{\bf C#1~}(19#2)~#3}
\def\prl#1#2#3{{\it Phys. Rev. Lett. }{\bf #1~}(19#2)~#3}
\def\rmp#1#2#3{{\it Rev. Mod. Phys. }{\bf #1~}(19#2)~#3}
\def\prep#1#2#3{{\it Phys. Rep. }{\bf #1~}(19#2)~#3}
\def\pr#1#2#3{{\it Phys. Rev. }{\bf D#1~}(19#2)~#3}
\def\np#1#2#3{{\it Nucl. Phys. }{\bf B#1~}(19#2)~#3}
\def\mpl#1#2#3{{\it Mod. Phys. Lett. }{\bf #1~}(19#2)~#3}
\def\arnps#1#2#3{{\it Annu. Rev. Nucl. Part. Sci. }{\bf
#1~}(19#2)~#3}
\def\sjnp#1#2#3{{\it Sov. J. Nucl. Phys. }{\bf #1~}(19#2)~#3}
\def\jetp#1#2#3{{\it JETP Lett. }{\bf #1~}(19#2)~#3}
\def\app#1#2#3{{\it Acta Phys. Polon. }{\bf #1~}(19#2)~#3}
\def\rnc#1#2#3{{\it Riv. Nuovo Cim. }{\bf #1~}(19#2)~#3}
\def\ap#1#2#3{{\it Ann. Phys. }{\bf #1~}(19#2)~#3}
\def\ptp#1#2#3{{\it Prog. Theor. Phys. }{\bf #1~}(19#2)~#3}

\vskip .2in

\noindent{\Large{\bf Figure Captions}}
\vskip .2in
\noindent{\bf Fig.~1:} {Two-loop Feynman diagrams contributing to
sfermion ($\tilde{f}$) masses. $\Phi$, $\bar \Phi$ and $\psi_{\Phi}$,
$\psi_{\bar \Phi}$ denote the messenger scalar and fermionic components
respectively. $\lambda$ is the gaugino and the wavy lines correspond to
gauge bosons.}
\vskip .2in
\noindent{\bf Fig.~2:} {Masses of the right-handed ($\tilde{e_R}$) and
left-handed ($\tilde{e_L}$) sleptons 
as a function of $\Lambda_D^2/\Lambda_S^2$. We have chosen a normalization
such that $m_{\tilde{e_R}}=100$ GeV when $\Lambda_D=0$.}

\vfill\eject

\begin{figure}
\hglue3.0cm
\epsfig{figure=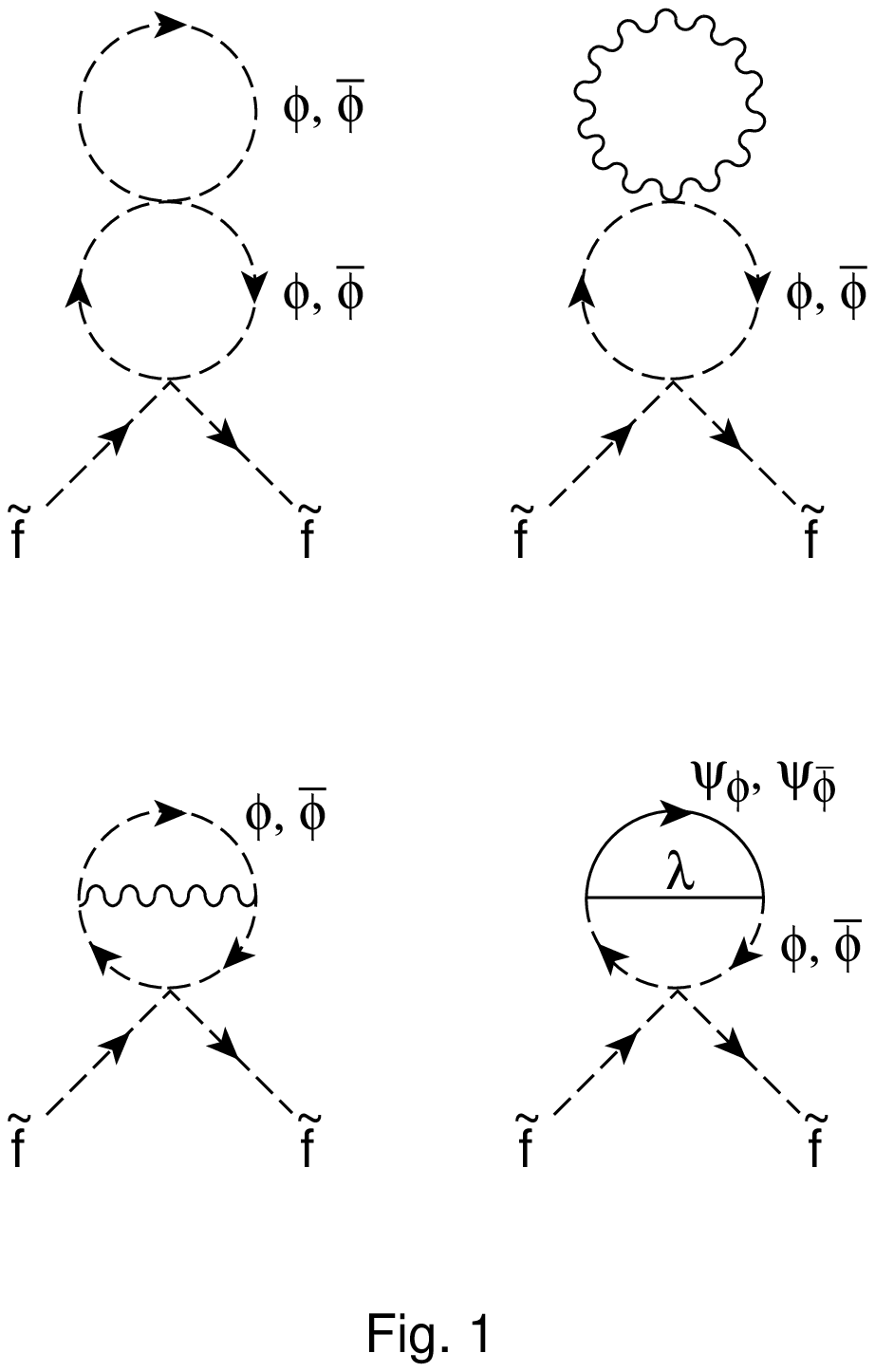,width=10cm}
\end{figure}

\begin{figure}
\hglue1.5cm
\epsfig{figure=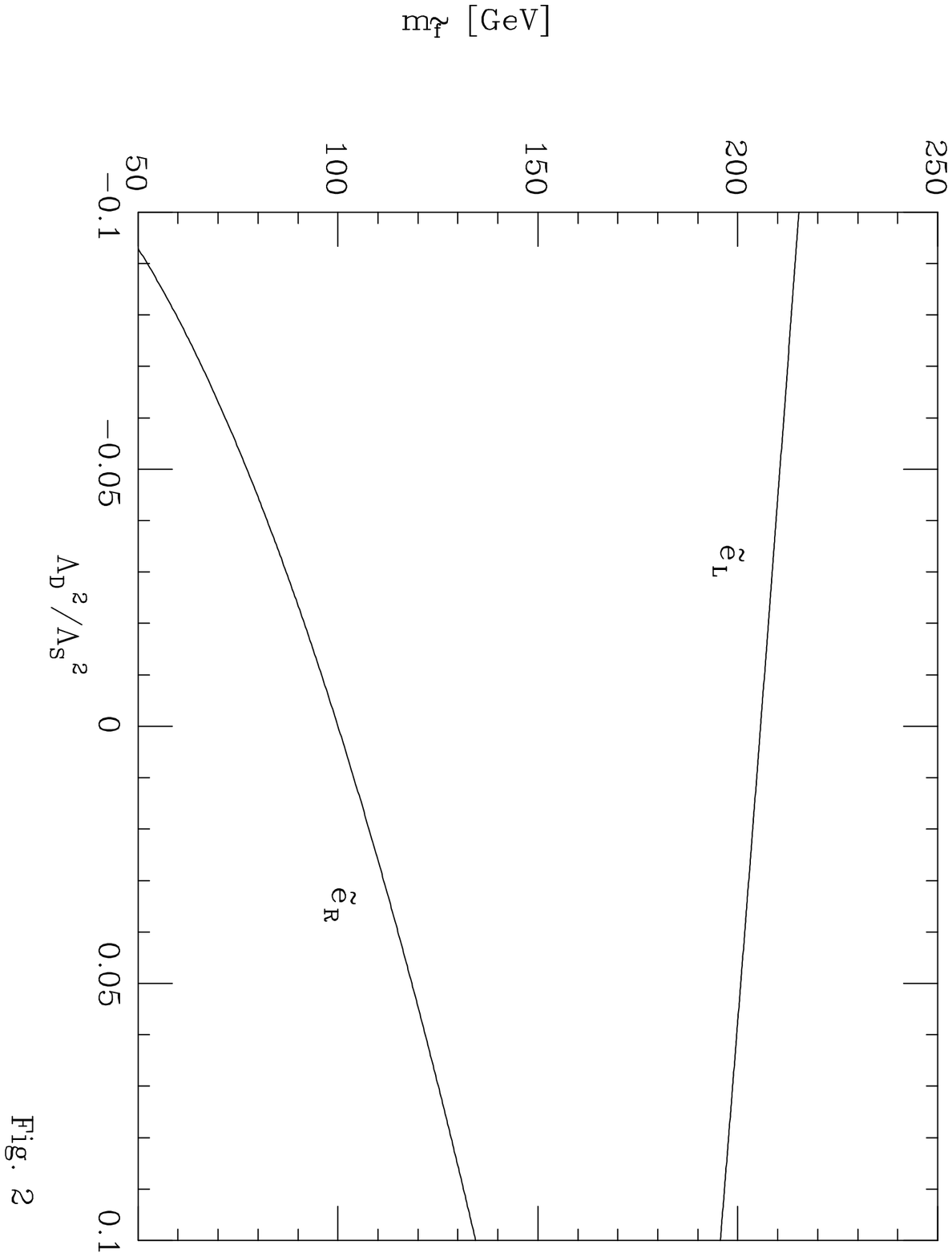,width=15cm}
\end{figure}

\end{document}